\documentclass[a4paper,12pt]{article}
\usepackage{amsthm,amsmath}
\usepackage{amssymb}
\usepackage{amsfonts}
\usepackage{color}
\usepackage[utf8]{inputenc}
\usepackage{hyperref}

\numberwithin{equation}{section}
\newtheorem{theorem}{Theorem}[section]

\newtheorem{lem}[theorem]{Lemma}
\newtheorem{remark}[theorem]{Remark}

\newcommand{\eps}{\varepsilon}

\newcommand{\n}{\mathbb{N}}
\newcommand{\R}{\mathbb{R}}

\newcommand{\e}{\rm e}
\def\beq{\begin{equation}}   \def\eeq{\end{equation}}
\def\bea{\begin{eqnarray}}  \def\eea{\end{eqnarray}}

\begin{document}
\title{When   Poisson and Moyal   Brackets are   equal?}

\date{}
\author{
Didier Robert\footnote{ Laboratoire de Mathématiques Jean Leray, Université de Nantes,
	2 rue de la Houssinière, BP 92208, 44322 Nantes Cedex 3, France\newline
\textit{Email: }{didier.robert@univ-nantes.fr}},
}
\maketitle

\begin{abstract}
   In the phase space $\R^{2d}$, let us denote $\{A,B\}$  the Poisson bracket of two smooth classical observables  and $\{A, B\}_\circledast $ their Moyal bracket,  defined as  the Weyl symbol
	  of $i[ \hat A,  \hat B]$, where $ \widehat A$ is the Weyl quantization of $A$ and  $[ \widehat A,  \widehat B]= \widehat A \widehat B- \widehat B \widehat A$ (commutator).\\
	In this note we prove that if  a given  smooth Hamiltonian $H$ on the phase space $\R^{2d}$,  with derivatives of moderate growth, satisfies 
	 $\{A,H\}= \{A, H\}_\circledast$ for any observable $A$ in the Schwartz space ${\mathcal S}(\R^{2d})$,  then, as it is expected,  $H$ must be  a polynomial of degree  at most 2 in $\R^{2d}$.
	\\
A  related  answer to this question is given in  the  Groenewold-van Hove Theorem \cite{Gotay, Groen, vHove}   concerning quantization of polynomial observables.
We consider here more general classes of  Hamiltonians.
	 
\end{abstract}

 \section{Introduction}	
 Let $H, A, B$ be  smooth classical observables on $\R^{2d}$ in the variables $X=(x,\xi)$.  The Poisson brackets
 is defined as $\{A,B\} =\partial_\xi A\cdot\partial_xB-\partial_xA\cdot\partial_\xi B$. So the classical  time evolution of $A$  determined by the Hamilton equation for $H$ is solution of the equation:
 \bea\label{cl}
 \frac {d}{dt}A(t) &=& \{A(t), H\} \\
 A(0) &=&A. \nonumber
 \eea
 The Weyl quantization $\widehat A$  of $A$  is defined as the following operator:
  \beq 
  \widehat A f(x) :=({\rm Op}^w_\hbar A) f(x)=(2\pi\hbar)^{-d} \int_{\R^{2d}} A\!\left(\frac{x+y}{2}, \xi\right) 
{\rm e}^{i\xi\cdot(x-y)/\hbar} f(y) \,dy\, d\xi
\eeq
for any $f\in{\mathcal S}(\R^d)$. Recall that  $f\in {\mathcal S}(\R^d)$ means that $f\in C^\infty(\R^d)$ and for any multiindex $\alpha,\beta$, 
$x^\alpha\partial_x^\beta f(x)$  is bounded on $\R^d$.\\
The quantum time evolution of the quantum observable $\widehat A$ must satisfy the Heisenberg equation
\bea\label{qu}
 \frac {d}{dt}\widehat A(t) &=& \frac{i}{\hbar}[ \widehat A(t),  \widehat H]\\
  \widehat A(0) &=&\widehat A
 \eea
 where $[\hat A, \hat B] = \hat A\hat B -\hat B\hat A$.\\
The Moyal bracket  of the observables $A, H$,  is defined such that 
\beq
\frac{i}{\hbar}[ \widehat A,  \widehat H]={\rm Op}^w_\hbar(\{A,H\}_\circledast).
\eeq
Notice that it results from   the Weyl quantization  calculus with a small parameter $\hbar$  that we have
$$
\lim_{\hbar \searrow 0}\{A,H\}_\circledast = \{A,H\}.
$$
{\em A natural question is to ask     when  the  classical dynamics generated by the Hamiltonian $H$ \eqref{cl} has an exact correspondence
with the quantum dynamics generated by $\hat H$ \eqref{qu}  (see below the quotation from Van Hove). In the correspondence principle stated by N. Bohr the Planck constant $\hbar$
 is supposed to be small.  The question  discussed here is for $\hbar>0$  fixed. }

  A well known trick to check the  correspondence Bohr principle  is to compute the time evolution of Gaussian coherent states. Let us denote
  $\varphi_Y =\hat T_Y\varphi_0$ the coherent state center in $Y\in\R^{2d}$ and $\varphi_0(x) = (\pi\hbar)^{-d/4}{\rm e}^{-\vert x\vert^2/2\hbar}$\\ ($\hat T_Y$ is  defined in the next section).
  We have \cite{CR} 
  $$
  \lim_{\hbar\searrow 0}\langle\varphi_Y,\hat A\varphi_Y\rangle = A(Y).
  $$
Hence taking the average of \eqref{qu} on $\varphi_Y$ and passing to the limit $\hbar\searrow 0$, we recover \eqref{cl}.

To define the Moyal bracket, there is a more explicite definition  by introducing the Moyal product  $A\circledast B$ (see the next section)
such that 
$$
({\rm Op}^w_\hbar A)({\rm Op}^w_\hbar B) = {\rm Op}^w_\hbar(A\circledast B).
$$
Then we have
$$
\{A,B\}_\circledast  = \frac{i}{\hbar}(A\circledast B-B\circledast A).
$$
These definitions make sense for $A, B\in {\mathcal S}(\R^{2d})$ and can be extended to suitable  classes of symbols with moderate growth. To be more explicite we introduce  the classes
${\mathbb{S}}_\delta^\mu$, for $\delta<1$, $\mu\in\R$. $A\in{\mathbb{S}}_\delta^\mu$ iff  $A\in C^\infty(\R^{2d})$ and   for any multiindex $\gamma\in\n^{2d}$
 we have:
$$
\vert\partial^\gamma_X A(X)\vert \leq C_\gamma\langle X\rangle^{\mu+\delta\vert\gamma\vert}
$$
Using Theorem A.1 in \cite{BR}, we can see that $A\circledast H$ is a smooth symbol if $H\in{\mathbb{S}}_\delta^\mu$ and $A\in{\mathbb{S}}_\delta^\nu$ where $\mu, \nu\in\R$ and $\delta<1/2$.
Our aim here is to prove the following result.
\begin{theorem}\label{MainThm} Assume that  $\hbar$  is fixed ($\hbar =1$). 
Let be $H\in {\mathbb{S}}_\delta^\mu$ for some $\mu\in\R$ and $\delta<1/2$.  Assume that  for any $A\in{\mathcal{S}}(\R^{2d})$  we have 
$ \{A, H\}_\circledast =  \{A, H\}$. Then $H(X)$ must be  a polynomial in $X=(x,\xi)$ of degree at most 2.  
\end{theorem}
\begin{remark}
It is well known that if $H$ is a polynomial of degree at most 2 then $\{A, H\}_\circledast =  \{A, H\}$  for any $A\in{\mathbb{S}}^\nu_0$. I do not know any  reference  for a proof of the converse statement. The proof given here is a consequence of basic properties of the Weyl quantization. 
\end{remark}
\begin{remark}
The usual proofs of the Groenewold-van Hove Theorem on the phase space $\R^{2d}$  concern  more general quantization procedures  but are restricted to polynomial symbols $A, H$.\\
       A quotation from \cite{vHove} p.66-67:\\
       "On \'etablit ensuite qu'une correspondance biunivoque entre grandeurs classiques et quantiques, ayant le  caract\`ere d'un isomorphisme entre alg\`ebres de Lie, existe entre les grandeurs repr\'esent\'ees par des polyn\^omes de degr\'e 0, 1, 2 en les variables $p_1,\cdots p_N,q_1,\cdots q_n$ mais ne peut \^etre \'etendue sans perdre ses propri\'et\'es essentielles \`a l'ensemble de toutes les grandeurs classiques"   \\
       The Theorem of Groenewold-van Hove is detailed  p.76 and the quadratic case p.87 of \cite{vHove}.\\
       Notice that the quadratic case is related with the metaplectic representation \cite{Gotay}.
\end{remark} 
\noindent
{\bf Acknowledgement} {\em In memory of Steve Zelditch who was at the origin of this question discussed with him twenty years ago.\\
I thank my colleagues Paul Alphonse and San Vu Ngoc for discussions concerning this question in June 2022.}

 \section{Weyl calculus}
 \subsection{Introduction to the Weyl quantization}
 In this section, we recall some  basic properties of the Weyl calculus (for more details see \cite{ho}).\\
 Weyl quantization starts by quantization of exponent of linear forms $L_Y(X) = \sigma(Y,X)=\eta\cdot x-y\cdot\xi$
  with $X=(x,\xi)$, $Y=(y,\eta)$. Apart the usual properties asked for an admissible quantization, Weyl quantization is uniquely determined by imposing
  that the Weyl symbol of ${\rm e}^{i\widehat{L_Y}}$ is ${\rm e}^{iL_Y}$. Recall that $ \widehat T(Y):={\rm e}^{-{i}\widehat{L_Y}}$   is the Weyl-Heisenberg translation operator by $Y$ in the phase space $\R^{2d}$. In other words the Weyl quantization $A\mapsto \hat A$ has to satisfy 
   ${\rm e}^{i\widehat{L_Y}} =  \widehat{({\rm e}^{iL_Y})}$.
  Then  for any observable $A$,  using a Fourier transform,  the  Weyl quantization $ A$ is defined  for any $\psi\in {\mathcal S}(\R^d)$, as
\beq\label{WT}
   \widehat A\psi = (2\pi)^{-d}\int_{\R^{2d}}{\tilde A}_\sigma(Y) \hat T(Y)\psi dY
\eeq 
  where ${\tilde A}_\sigma(Y)=\int_{\R^{2d}} A(z){\rm e}^{-i\sigma(Y,z)}dz$ is the symplectic Fourier transform of $A$ (in the sense of distributions).
   So that the family $\{T(Y)\}_{Y\in\R^{2d}}$ is an over-complete   basis for operators   between the Schwartz spaces ${\mathcal S}(\R^d)$ and 
   ${\mathcal S}'(\R^d)$. $\tilde A_\sigma$ is the covariant symbol of $\hat A$ and $A$ the contravariant symbol of $\hat A$.
   
  \begin{remark}
  Notice that from \eqref{WT} for any symbol $A$ and for any linear form $L_Z$ we get
\beq\label{CL}
   i[\hat A,\hat L_Z] = \widehat{\{A, L_Z\}}.
  \eeq
  It is enough to prove \eqref{CL} for $\hat A=\hat T(Y)$. This is done using the translation property of the Heisenberg unitary operators  $\hat T(Y)$
   where $Y=(y,\eta)$, $D_x =i^{-1}\nabla_x$,  we have:
  \beq\label {evol1}
  \hat T(sY)^*\begin{pmatrix}x\\D_x\end{pmatrix}  \hat T(sY) = \begin{pmatrix}x-sy\\D_x-s\eta\end{pmatrix},\;s\in\R,
\eeq
  $s\mapsto (x-sy, \xi-s\eta)$ is the classical  translation motion for the linear Hamiltonian $L(Y)$. \\
  For quadratic Hamiltonians the classical flow is a time dependent linear symplectic map and  the extension of \eqref{CL} and \eqref{evol1} to quadratic Hamiltonians can be proved by the same method \cite[Theorem 15, p.65]{CR}.
  \end{remark}
 \subsection{The Moyal Product}
We first recall the formal product rule for quantum observables 
with Weyl 
quantization. Let $A, B \in {\mathcal S}(\R^{2d})$. The Moyal product $C:=A\circledast B$ is the  
observable 
$C$ such that $\widehat{A}\cdot\widehat{B} = \widehat{C}$. Some computations with 
the Fourier 
transform give the following well known formulas \cite{ho} (see also \cite{robook})
\beq\label{moy1}
(A\circledast B)(X) = (\pi\hbar)^{-2d}\int\!\!\int_{\R^{2d}\times\R^{2d}}{\rm e}^{\frac{2i}{\hbar}\sigma(u,v)}A(X+u)B(X+v)dudv.
\eeq
Some more computations with 
the Fourier  transform give the following  formula :
\beq\label{moy2}
(A\circledast B)(x,\xi) = 
\exp\left(\frac{i\hbar}{2}\sigma(D_q,D_p;D_{q^\prime},D_{p^\prime})\right)A(q,p) 
B(q^\prime,p^\prime)\vert_{(q,p)=(q^\prime,p^\prime)=(x,\xi)}, 
\eeq 
where $\sigma$ is the symplectic bilinear form $\sigma((q,p), (q',p'))=p\cdot q'-p'\cdot q$ and 
$D=i^{-1}\hbar\nabla$. 
By expanding the exponential term in a formal power series in $\hbar$ we get 
\beq\label{prod2} 
C(x,\xi) = 
\sum_{j\geq 
0}\frac{\hbar^j}{j!}\left(\frac{i}{2}\sigma(D_q,D_p;D_{q^\prime},D_{p^\prime}) \right)^j 
A(q,p)B(q^\prime,p^\prime)\vert_{(q,p)=(q^\prime,p^\prime)=(x,\xi)}. 
\eeq 
So that $C(x,\xi)$ is a formal power series in $\hbar$ with coefficients 
given by 
\beq \label{exprod}
C_j(A,B;x,\xi) = \frac{1}{2^j}\sum_{\vert\alpha+\beta\vert=j} 
\frac{(-1)^{\vert \beta\vert}}{\alpha!\beta!} 
(D^\beta_x\partial^\alpha_\xi A).( D^\alpha_x\partial^\beta_\xi B)(x,\xi). 
\eeq 
%It is  enough for our purpose to assume that $A, B$ grow polynomialy at infinity.
%\end{document}
\medskip 
Furthermore we need  a remainder estimates for the expansion of the Moyal product.\\
For every $N \geq 1$. we denote
\beq 
R_N(A,B;X):=
A\circledast B(X) - \sum_{0\leq j \leq N}\hbar^jC_j(X) .
\eeq 
The following estimate is a particular case of Theorem A.1 in \cite{BR} see also Remark A.3.

\begin{lem}\label{lem:Moyalest}
Let be $A\in{\mathbb S}^{\mu_A}_\delta$ and  $B\in{\mathbb S}^{\mu_B}_\delta$, $\delta<1/2$,   then for any $N\geq 1$, $\gamma\in\n^{2d}$, $M\geq M_0$  there exists $C_{N,\gamma,M}>0$(independent of $(A, B)$)
such that 
\bea\label{remest1} 
&\vert\partial_X^\gamma R_N(A,B;X)\vert \leq C_{N,\gamma,M}\hbar^{N+1}\sum_{\substack{\vert\alpha+\beta\vert=N+1\\ \vert\mu+\nu\vert\leq M+\vert\gamma\vert}}\\\nonumber&\sup_{u, v\in\R^{2d}} (1+\vert u\vert^2+\vert v\vert^2)^{(M_0-M)/2}\vert\partial_u^{(\alpha,\beta)+\mu}A(X+u)\vert\vert\partial_v^{(\beta,\alpha)+\nu}B(X+v)\vert
\eea
In particular $ R_N(A,B,X)\in \mathbb S_\delta^{\mu_{AB}}$ for some $\mu_{AB}\geq \mu_A+\mu_B$.
\end{lem} 
For proving  this Lemma  one assume first that $A, B\in{\mathcal{S}}(\R^{2d})$. For the general case we put $A_\varepsilon(X)={\rm e}^{-\epsilon\vert X\vert^2}A(X)$, 
$B_\varepsilon(X)={\rm e}^{-\epsilon\vert X\vert^2}B(X)$ and pass to the limit for $\varepsilon\searrow 0$. In the appendix we give more details.
We also need to use the following Lemma.
\begin{lem}\label{lim}
 $A\in{\mathbb S}^{\mu_A}_\delta$ and  $B\in{\mathbb S}^{\mu_B}_\delta$, $\delta<1/2$. Then uniformly in every compact of $\R_X^{2d}$, we have
 $$
 \lim_{\epsilon\searrow 0}(A_\epsilon\circledast B)(X) =  \lim_{\epsilon\searrow 0}(A\circledast B_\epsilon)(X) = (A\circledast B)(X).
 $$
In particular we have
$$
 \lim_{\epsilon\searrow 0}\{A_\epsilon, B\}_\circledast (X) = \{A, B\}_\circledast (X).
 $$
\end{lem}
For completeness a proof is given in the appendix B.
\section{Proof of Theorem(\ref{MainThm})}
 Here $\hbar=1$.  Notice first that from Lemma \ref{lim} we also have for any $A\in \mathbb{S}_0^0$,
$ \{A,H\}_\circledast =   \{A,H\}$. 
 So it is enough to consider the test observables \\$A=T_Y:={\rm e}^{-iL_Y}$ ($Y\in\R^{2d}$). \\ 
 We have 
 $$
 \widehat T_Y\widehat H\widehat T_Y^* = [\widehat T_Y, \widehat H]\widehat T_Y^* + \widehat H
 $$
Using the assumption  of  Theorem(\ref{MainThm})    and   Lemma \ref{lim}  we get 
\beq\label{MPC1}
\frac{1}{i}(\{T_T^*, H\}\circledast T_Y)(X) = H(X+Y)-H(X), \forall X, Y\in\R^{2d}.
\eeq
Computing the Poisson bracket in \eqref{MPC1} gives
\beq\label{MPC2}
(((y\cdot\partial_x H +\eta\cdot\partial_\xi H)T_Y^*)\circledast T_Y)(X) = H(X+Y) -H(X),\;\forall X, Y\in\R^{2d}.
\eeq
Our aim is to prove that \eqref{MPC2}  implies that $H(X)$ is a polynomial of degree at most 2. For that purpose we shall compute 
the asymptotic expansion  as $Y\rightarrow 0$ of the left hand side  of \eqref{MPC2} and compare it with the Taylor expansion for  $H(X+Y)$ modulo
 $O(\vert Y\vert^4)$. From that we shall conclude that  all the third  order derivatives of $H$ vanish for $X$ in any bounded subset of $\R^{2d}$ hence the conclusion will follow.
 
 We  have  
 $$
 \partial_x^\alpha\partial_\xi^\beta T_Y = i^{-\vert\alpha+\beta\vert}\eta^\alpha y^\beta T_Y
 $$ 
 Let us denote by $C(X,Y)$ the left hand side in \eqref{MPC2}. 
 So using Lemma(\ref{lem:Moyalest}) uniformly in every compact in $X\in\R^{2d}$,   we have 
 $$
 C(X,Y) = \sum_{0\leq j\leq 2}(C_j(X,Y) + O(\vert Y\vert^4),
 $$
 where 
 \bea
 C_0(X,Y) &= & Y\cdot\nabla_XH(X)\\
 C_1(X,Y) &=& \frac{1}{2}Y\cdot \nabla_X^2H(X)Y,
 \eea
 where $\nabla_X^2H(X)$ is the Hessian matrix of $H$.\\
 Let us compute now $C_2(X,Y)$, which is   an homogeneous polynomial of degree 3 in $Y$.\\
 For simplicity let us consider the 1-D case. The same computation can clearly  be done for $d>1$.\\
 Using \eqref{prod2} we get with $Y=(y,\eta)$, 
 \beq\label{C2}
 C_2(X,Y) = \frac{1}{8}\left(y^3\partial_x^3H +\eta^3\partial_\eta^3H -y^2\eta\partial_\xi\partial_x^2H - y\eta^2\partial_\xi^2\partial_xH\right).
 \eeq
% \end{document}
 According \eqref{MPC2}, $C_2(X,Y)$ must  coincide with the term of order 3  in $Y$ of the Taylor expansion in $X$ for $H(X+Y)-H(X)$.
 But this is possible only if $\partial_x^3H=\partial_\eta^3H =\partial_\xi\partial_x^2H =\partial_\xi^2\partial_xH=0$   for any $(x,\xi)\in \R^2$.
 So $H$ must be a polynomial of degree $\leq 2$. $\square$
 
 \section{Extension  to polynomials of arbitrary degree }
 The asymptotic expansion in $\hbar$ in the Moyal product suggests  to introduce the following semi-classical approximations of the Moyal bracket:
 $$
 \{ A,B\}_{\circledast,m} = \{A,B\} +\hbar^2\{A,B\}_3 +\cdots +\hbar^{2m}\{A,B\}_{2m+1},
 $$
 where $\{A,B\}_j = \frac{i}{\hbar}(C_j(A,B)-C_j(B,A))$  (notation of  \eqref{exprod}). 
 Notice that $\{A,B\}_j =0$ for $j$ even.\\
 It is clear that if $H$ is a polynomial of degree at most $2m+2$ then we have $ \{ A,H\}_{\circledast,m} = \{ A,H\}_{\circledast}$  
 for any $A$.  Conversely we have
\begin{theorem}\label{thm:orderm} Assume $\hbar =1$  and $H\in \mathbb{S}^\mu_\delta$, $\mu\in \R$, $\delta<1/2$. If for any $A\in {\mathcal S}(\R^{2d})$ we have $ \{ A,H\}_{\circledast,m} = \{ A,H\}_{\circledast}$
 then $H$  must be  a polynomial of degree at most $2m+2$.
 \end{theorem}
 {\em Proof}. Here we give a proof different from the case $m=0$, without connection with the Taylor   formula,  for simpler computations.\\
 Using Lemma \ref{lem:Moyalest} we have, uniformly in every compact in $X\in\R^{2d}$,  
 \beq\label{comp1}
 T_Y^*\left( \{T_Y,H\}_{\circledast}(X) - \{T_Y,H\}_{\circledast,m}(X) \right)= {\mathcal O}(\vert Y\vert^{2m+3}),\; Y\rightarrow 0.
 \eeq
 Moreover from \eqref{exprod} we get:
 \beq\label{comp2} 
  T_Y^*\{T_Y,H\}_{2j+1}(X) =\frac{1}{2^{j+1}}\sum_{\vert\alpha+\beta=2j+1}\frac{y^\alpha\eta^\beta}{\alpha!\beta!}\partial_x^\alpha\partial_\xi^\beta H(X).  
   \eeq
Using the assumption of Theorem \ref{thm:orderm}. and \eqref{comp1}  we get that 
$$
 T_Y^*\{T_Y,H\}_{2m+3}(X) = {\mathcal O}_X(\vert Y\vert^{2m+5}).
 $$
 But  $T_Y^*\{T_Y,H\}_{2m+3} $ is an homogeneous polynomial of degree $2m+3$ in $Y$ so we get that this polynomial is 0 and from
 \eqref{comp2} we get that $ \partial_x^\alpha\partial_\xi^\beta H(X) =0 $   for $\vert\alpha+\beta\vert=2m+3$. Then we can conclude that $H(X)$ is a polynomial of degree at most $2m+2$ in $X\in\R^{2d}$. $\square$.
 \appendix
 \section{Proofs for formula \eqref{moy1} and \eqref{moy2}}
 It is enough to assume that $A, B\in{\mathcal S}(\R^{2d})$.\\
 Recall first the relationship between Weyl symbols 	and integral kernel of $\hat A$.
 We have 
 $$
 K_{\hat A}(x,y) = (2\pi\hbar)^{-d}\int_{\R^{d}}{\rm e}^{\frac{i}{\hbar}(x-y)\cdot\eta}A(\frac{x+y}{2}, \eta)d\eta
 $$
 and 
 $$
 A(x,\xi) = \int_{\R^d}{\rm e}^{-\frac{i}{\hbar}\xi\cdot t}K_{\hat A}(x+t/2, x-t/2)dt.
 $$
 Using these formulas and the relation $K_{\hat A\hat B}(x,z) = \int_{\R^d}K_{\hat A}(x,y)K_{\hat B}(y,z)dy$ we get
 \bea
 (A\circledast B)(X) = &(2\pi\hbar)^{-2d}\int_{\R^{4d}} \exp\left(\frac{i}{\hbar}(-t\cdot\xi +(x-y-t/2)\cdot\eta +(y-x+t/2))\cdot\zeta\right).\nonumber  \\
 &.A((x+y)/2+t/4,\eta)B((x+y)/2-t/4,\zeta)d\zeta d\eta dydt
 \eea
 Then after a change of variables  in the integral $v_\xi = \zeta -\xi, u_\xi = \eta-\xi$, $u_x = (y-x)/2 +t/4, v_x= (y-x)/2 -t/4$,  we get formula \eqref{moy1}, with $u=(u_x,u_\xi), \;v=(v_x,v_\xi)$, 
 $$
   (A\circledast B)(X) =(\pi\hbar)^{-2d}\int_{\R^{2d}\times\R^{2d}}{\rm e}^{\frac{2i}{\hbar}\sigma(u,v)}A(X+u)B(X+v)dudv.
   $$
   To get formula \eqref{moy2} we notice that $(u,v)\mapsto 2\sigma(u,v)$ is non degenerate and its matrix is $G:=\begin{pmatrix} 0 & -J\\J & 0\end{pmatrix}$,
    so $G^{-1} =G$. Hence using Fourier transform in $(u, v)$ and  the Fourier multiplier  formula we get \eqref{moy2}.$\square$
 \section{Proofs for  Lemmas \ref{lem:Moyalest} and  \ref{lim}}
 \subsection{Proof of lemma \ref{lim}}
 Using   \eqref{moy1}  for $A_\epsilon\circledast B$ we split the integral in two pieces  : \\ $1 =\chi_0(\vert u\vert^2 +\vert v\vert^2) + \chi_1(\vert u\vert^2 +\vert v\vert^2)$,
  where  $\chi_0\in C_0^\infty(\R)$, $\chi_0(t)=1$ for $\vert t\vert \leq 1/2$.  On the support of $\chi_0$ we can obviously pass to the limit in $\epsilon$. On the support of $\chi_1$ we first perform integrations by parts with the   differential operator $L$  several times to get a uniformly and absolutely convergent integral, 
  %%%%%%%%%%%
  \[
L = \frac{Ju\cdot\partial_v-Jv\cdot\partial_u}{|u|^2+|v|^2},
\]
using that $L \e^{\frac{2i}{\hbar}\sigma(u,v)} = L \e^{\frac{2i}{\hbar}\sigma(u,v)} = \frac{2i}{\hbar}\e^{\frac{2i}{\hbar}\sigma(u,v)}$. 
On the support of $\chi_1$, performing $4d+1$ integrations by parts for gaining enough decay to ensure integrability in $(u,v)\in\R^{4d}$. 
  Then passing to the limit in $\epsilon$ we get 
  $
 \lim_{\epsilon\searrow 0}(A_\epsilon\circledast B)(X) = (A\circledast B)(X)$ and the same for $
 \lim_{\epsilon\searrow 0}(A\circledast B_\varepsilon)(X) = \{A, B\}_\circledast (X)$.\\
 The other properties follow.
 $\square$
 %\end{document}
 %%%%%%%%%%%%%%%
 %%%%%%%%%%%%%
 \subsection{Proof of Lemma \ref{lem:Moyalest}}
 From \eqref{moy1}, by Fourier transform computations and application of the Taylor formula, we get the following formula for the remainder,
\beq\label{reprem} 
R_N(A,B,X) = \frac{1}{N!}\left(\frac{i\hbar}{2}\right)^{N+1} 
\int_0^1(1-t)^NR_{N,t}(X;\hbar)dt,
\eeq
where
\begin{align*}
&R_{N,t}(X;\eps) = \\
&(2\pi\bar t)^{-2d} 
\int\int_{\R^{2d}\times \R^{2d}} 
\exp\left(-\frac{i}{2t\hbar}\sigma(u,v)\right)\sigma^{N+1}(D_u,D_v) 
A(u+X)B(v+X)dudv. 
\end{align*}
Notice that the integral is an oscillating integral as we shall see below.  So we shall use the following lemma :
\begin{lem}\label{foi} 
There exists a constant $C_d>0$ such that for any $F\in{\mathcal S}(\R^{2d}\times \R^{2d})$ 
the integral 
\beq
I(\lambda) = \lambda^{2d}\int \int_{\R^{2d}\times \R^{2d}}\exp[-i\lambda 
\sigma(u, v)]F(u,v)dudv. 
\eeq 
satisfies  the following estimate:\\
  for any $M>0$ there exists $ C_M>0$, independent of $F$,  such that

\beq
\vert I(\lambda\vert \leq C_{M} \sup_{\stackrel{u,v\in\R^{2d}}{\vert\alpha +\beta\vert\leq M+4d+1}}(1+\vert u\vert^2 +\vert v\vert^2)^{(4d+1-M)/2}\vert\partial_u^\alpha\partial_v^\beta F(u,v)\vert
\eeq
\end{lem} 
\noindent A proof  will be given later.\\
Using this Lemma  for $A, B\in{\mathcal S}(\R^{2d})$ with the integrand 
\[
F_{N,\gamma}(X;u,v) = \pi^{-2d} \ \partial_X^\gamma\left(\sigma^{N+1}(D_u,D_v) A(u+X)B(v+X)\right)
\] 
and the parameter $\lambda=1/(2t\hbar)$. We then have that 
\[
|\partial_X^\gamma R_{N,t}(X;\eps)| \le C_d \sup_{u,v\in\R^{2d}\atop |\alpha|+|\beta|\le 4d+1}|\partial^\alpha_u\partial^\beta_v F_{N,\gamma}(X;u,v)|.
\]
Moreover, we have the elementary estimate
\begin{align} \label{AI2}
\vert \sigma^{N+1}(D_u,D_v)A(u)B(v)\vert \leq 
(2d)^{N+1}\sup_{\vert \alpha\vert+\vert \beta\vert=N+1} 
\vert\partial_x^\alpha\partial_\xi^\beta 
A(x,\xi)\partial_y^\beta\partial_\eta^\alpha B(y,\eta)\vert. 
\end{align}
Together with the Leibniz formula, we then get the claimed result with universal constants.
For symbols $A\in{\mathbb S}_0^\mu$ and $B\in{\mathbb S}_0^\nu$ we argue by localisation. We use 
$A_\epsilon(u) = {\rm e}^{-\epsilon u^2}A(u)$ and 
$B_\epsilon(v) = {\rm e}^{-\epsilon v^2}B(v)$ for $\epsilon
>0$ and 
pass to the limit as  $\epsilon \rightarrow 0$.

\subsection{Proof of the Lemma \ref{foi}}

 We consider the same  cut-off $\chi_0$ as above.
 We split $I(\lambda)$ into 
two pieces and write 
$I(\lambda)=I_0(\lambda) + I_1(\lambda) $ with 
\begin{align*}
I_0(\lambda)& = \lambda^{2d}\int\!\!\int_{\R^{2d}\times \R^{2d}}\exp[-i\lambda 
\sigma(u, v)] 
\chi_0((u^2+v^2))F(u,v)dudv, \\ 
I_1(\lambda) &= \lambda^{2d}\int\!\!\int_{\R^{2d}\times \R^{2d}}\exp[-i\lambda 
\sigma(u, v)] 
(1-\chi_0)(u^2+v^2))F(u,v)dudv.
\end{align*}
We notice that $(u,v)\mapsto \sigma(u,u)$ is a quadratic  non-degenerate real  form on $\R^{4d}$.\\
Let us estimate  $I_1(\lambda)$. We can   integrate by parts with the 
differential operator 
\[
L = \frac{i}{|u|^2+|v|^2}\left(Ju\cdot\frac{\partial}{\partial 
v}-Jv\cdot\frac{\partial}{\partial u}\right),
\]
using that $L \e^{-i\lambda\sigma(u,v)} = L \e^{-i\lambda Ju\cdot v} = \lambda\e^{-i\lambda\sigma(u,v)}$. 
For $I_1(\lambda)$, the integrand is supported outside  the ball of radius $1/\sqrt2$ in $\R^{4d}$. Performing $4d+1$ integrations by parts for gaining enough decay to ensure integrability in $(u,v)\in\R^{4d}$, we get a constant $c_d$ such that 
\beq\label{AI2} 
\vert I_1(\lambda)\vert \leq c_d
\sup_{u,v \in \R^{2d}\atop \vert \mu\vert+\vert \nu\vert\leq 4d+1} 
\vert\partial^\mu_u \partial^\nu_vF(u,v)\vert. 
\eeq 
But we need to control the  behaviour for $u^2 +v^2$ large, so with $M$ more integrations by parts we get
\beq\label{I1}
\vert I_1(\lambda)\vert \leq C_{M} \sup_{\stackrel{u,v\in\R^{2d}}{\vert\alpha +\beta\vert\leq M}}(1+\vert u\vert^2 +\vert v\vert^2)^{(4d+1-M)/2}\vert\partial_u^\alpha\partial_v^\beta F(u,v)\vert
\eeq

To estimate $I_0(\lambda)$ we apply the stationary phase. 
The symmetric matrix of the quadratic form $\sigma(u,v)$ is 
$$
A_\sigma = \begin{pmatrix}0 & -J\\
J & 0\end{pmatrix}.
$$
So the stationary  phase theorem (\cite{ho}, Vol.I, section 7.7), noticing that the leading term in  the stationary phase theorem is of order $\lambda^{-2d} $,  we get 
\beq
\vert I_0(\lambda)\vert \leq c'_d
\sup_{u,v \in \R^{2d}\atop \vert \mu\vert+\vert \nu\vert\leq 2d+3} 
\vert\partial^\mu_u \partial^\nu_vF(u,v)\vert. 
\eeq 
%\end{document}

\end{document}